\documentclass[preprint2,longabstract]{aastex}

\newcommand{\degrees}[1]{\ensuremath{#1^\circ}}

\shorttitle{Six Pulsars Detected by the {\em Fermi} LAT}
\shortauthors{Weltevrede et al.}

\begin{document}

\title{Gamma-ray and Radio Properties of Six Pulsars Detected by the {\em Fermi} Large Area Telescope}


\author{
P.~Weltevrede\altaffilmark{2,3,1}, 
A.~A.~Abdo\altaffilmark{4,5}, 
M.~Ackermann\altaffilmark{6}, 
M.~Ajello\altaffilmark{6}, 
M.~Axelsson\altaffilmark{7,8}, 
L.~Baldini\altaffilmark{9}, 
J.~Ballet\altaffilmark{10}, 
G.~Barbiellini\altaffilmark{11,12}, 
D.~Bastieri\altaffilmark{13,14}, 
B.~M.~Baughman\altaffilmark{15}
% ...
}
\altaffiltext{1}{Corresponding authors: S.~Johnston, Simon.Johnston@atnf.csiro.au; D.~A.~Smith, smith@cenbg.in2p3.fr; P.~Weltevrede, patrick.weltevrede@manchester.ac.uk.}
\altaffiltext{2}{Australia Telescope National Facility, CSIRO, Epping NSW 1710, Australia}
\altaffiltext{3}{Jodrell Bank Centre for Astrophysics, School of Physics and Astronomy, The University of Manchester, M13 9PL, UK}
\altaffiltext{4}{Space Science Division, Naval Research Laboratory, Washington, DC 20375, USA}
\altaffiltext{5}{National Research Council Research Associate, National Academy of Sciences, Washington, DC 20001, USA}
\altaffiltext{6}{W. W. Hansen Experimental Physics Laboratory, Kavli Institute for Particle Astrophysics and Cosmology, Department of Physics and SLAC National Accelerator Laboratory, Stanford University, Stanford, CA 94305, USA}
\altaffiltext{7}{Department of Astronomy, Stockholm University, SE-106 91 Stockholm, Sweden}
\altaffiltext{8}{The Oskar Klein Centre for Cosmoparticle Physics, AlbaNova, SE-106 91 Stockholm, Sweden}
\altaffiltext{9}{Istituto Nazionale di Fisica Nucleare, Sezione di Pisa, I-56127 Pisa, Italy}
\altaffiltext{10}{Laboratoire AIM, CEA-IRFU/CNRS/Universit\'e Paris Diderot, Service d'Astrophysique, CEA Saclay, 91191 Gif sur Yvette, France}
\altaffiltext{11}{Istituto Nazionale di Fisica Nucleare, Sezione di Trieste, I-34127 Trieste, Italy}
\altaffiltext{12}{Dipartimento di Fisica, Universit\`a di Trieste, I-34127 Trieste, Italy}
\altaffiltext{13}{Istituto Nazionale di Fisica Nucleare, Sezione di Padova, I-35131 Padova, Italy}
\altaffiltext{14}{Dipartimento di Fisica ``G. Galilei", Universit\`a di Padova, I-35131 Padova, Italy}
\altaffiltext{15}{Department of Physics, Center for Cosmology and Astro-Particle Physics, The Ohio State University, Columbus, OH 43210, USA}
% ...

\begin{abstract}
We report the detection of pulsed $\gamma$-rays for PSRs J0631+1036,
J0659+1414, J0742--2822, J1420--6048, J1509--5850 and J1718--3825
using the Large Area Telescope (LAT) on board the {\em Fermi Gamma-ray
Space Telescope} (formerly known as GLAST). Although these six pulsars
are diverse in terms of their spin parameters, they share an important
feature: their $\gamma$-ray light curves are (at least given the
current count statistics) single peaked. For two pulsars
there are hints for a double-peaked structure in the light curves. The
shapes of the observed light curves of this group of pulsars are
discussed in the light of models for which the emission originates
from high up in the magnetosphere.  The observed phases of the
$\gamma$-ray light curves are, in general, consistent with those
predicted by high-altitude models, although we speculate that the
$\gamma$-ray emission of PSR J0659+1414, possibly featuring the
softest spectrum of all {\em Fermi} pulsars coupled with a very low
efficiency, arises from relatively low down in the
magnetosphere. High-quality radio polarization data are available
showing that all but one have a high degree of linear polarization.
This allows us to place some constraints on the viewing geometry and
aids the comparison of the $\gamma$-ray light curves with high-energy
beam models.
\end{abstract}


\section*{}
\clearpage

\section{Introduction}

\newlength{\figwidth}
\setlength{\figwidth}{0.95\hsize}

\begin{table*}[t]
\caption{\label{RefTableSpin}Rotational and derived
parameters for six pulsars.
}
\begin{center}
\begin{tabular}{llccccc}
\hline
\hline
\multicolumn{1}{c}{PSR} & \multicolumn{1}{c}{PSR}  & $P$ & $\dot{E}$ & $\tau_\mathrm{c}$\tablenotemark{a} & $B_\mathrm{S}$\tablenotemark{b} & $B_\mathrm{LC}$\tablenotemark{c}\\
\multicolumn{1}{c}{(J2000)} & \multicolumn{1}{c}{(B1950)} & (sec) & ($10^{35}$ erg\,s$^{-1}$) & ($10^{3}$ yr) & ($10^{12}$ G) & ($10^{3}$ G)\\
\hline
J0631+1036  &           &  0.288 &   1.73 &  43.6 & $5.55$ & $2.18$\\
J0659+1414  & B0656+14  &  0.385 &   0.38 & 111   & $4.66$ & $0.77$\\
J0742--2822 & B0740--28 &  0.167 &   1.43 & 157   & $1.69$ & $3.43$\\
J1420--6048 &           &  0.068 & 104    &  13.0 & $2.41$ & $71.3$\\
J1509--5850 &           &  0.089 & 5.15   & 154   & $9.14$ & $12.2$\\
J1718--3825 &           &  0.075 & 12.5   &  89.5 & $1.01$ & $22.6$\\
\hline
\end{tabular}
\end{center}
\tablenotetext{a}{Spin-down age $\tau_\mathrm{c}=P/(2\dot{P})$.}
\tablenotetext{b}{Magnetic field strength at the surface of the star
in Gauss $B_\mathrm{S}=3.2\times10^{19}(P\dot{P})^{1/2}$.}
\tablenotetext{c}{Magnetic field strength at the light cylinder in
Gauss $B_\mathrm{LC}=3.0\times10^{8}(\dot{P}/P^5)^{1/2}$.}
\end{table*}

The {\em Fermi Gamma-ray Space Telescope} (formerly known as GLAST)
was successfully launched on 2008 June 11. The study and discovery of
$\gamma$-ray pulsars is one of the major goals of this
mission. Studying pulsars at these high energies is important, because
a large fraction of the total available spin-down energy loss rate
($\dot{E}=4\pi^2I\dot{P}P^{-3}$) is emitted in $\gamma$-rays. Here $I$
is the moment of inertia of the star (generally taken to be
$10^{45}$~g$\,$cm$^{2}$), $P$ its spin period and $\dot{P}$ its spin
down rate. By studying individual {\em Fermi} detections as well as
the population of $\gamma$-ray pulsars as a whole, models for the
high-energy emission can be constrained (e.g. \citealt{hgg07,wrwj09}).

The models can be divided into three different families that place the
emitting regions at different locations in the pulsar
magnetosphere. In the so-called polar-cap model (e.g. \citealt{dh96})
the $\gamma$-ray photons are produced close to the neutron star
surface (within a few stellar radii) near the magnetic axis.  At the
other extreme are the outer-gap models (e.g. \citealt{mor83,chr86a,ry95}),
which place the emitting region near the light cylinder. Finally, in
slot-gap models (e.g. \citealt{mh04}) the particle acceleration occurs
in a region bordering the last open field lines at a large range of
emission altitudes. The two-pole caustic model (\citealt{dr03}) is a
geometrical realization of the slot-gap model.

It has recently been demonstrated that $\gamma$-ray pulsars can be
discovered via blind searches in the {\em Fermi} data
\citep{aaa+09c}. Nevertheless the detection threshold for
pulsed $\gamma$-rays is lower when accurate positions and spin
frequencies (including their unpredictable timing irregularities) are
already known. Therefore a set of pulsars with $\dot{E}>10^{34}$
erg\,s$^{-1}$ is being timed in the radio band, allowing {\em Fermi}
to search for pulsations with the highest possible sensitivity
\citep{sgc+08}. In addition these radio observations allow us to
determine the difference in arrival time of the $\gamma$-ray pulses
with respect to the radio pulses, an important parameter to
distinguish between different high-energy models.

In this paper we report the detection of pulsed $\gamma$-rays for six
pulsars that were found by folding the {\em Fermi} data on the
ephemerides obtained from radio observations
(e.g. \citealt{wjm09}). These {\em Fermi} detections will be
included (although with more limited count statistics) in the LAT
pulsar catalog paper \citep{aaa+09e}.
The six pulsars of this paper have moderate to large
spin-down luminosities ($\dot{E} > 10^{34.5}$ erg\,s$^{-1}$, see Table
\ref{RefTableSpin}) and have (at least given the current count
statistics)
$\gamma$-ray light
curves which are consistent with a single peak. Five of the six
pulsars have a strong degree of linear polarization in the radio band,
which can be used to constrain the emission geometry. Therefore the
combination of radio and $\gamma$-ray data for these objects make them
valuable for tests of the underlying beaming geometry.

The paper is organised such that we will start with describing
the {\em Fermi} observations in Section \ref{SctObs}. This is followed
by a description of the methods used to constrain the emission
geometry using radio data in Section \ref{SctMethods}.  The results,
including the $\gamma$-ray light curves and spectral parameters
obtained with {\em Fermi}, are presented in Section \ref{SctResults}
and discussed in Section \ref{SctDiscussion}. Finally the results and
conclusions are summarised in Section \ref{SctConclusions}.

\section{Observations}
\label{SctObs}

\begin{table*}[t]
\caption{\label{tableGammaLightcurves}Parameters of the analysis of
the $\gamma$-ray light curves. 
}
\begin{center}
\begin{tabular}{lcrlc}
\hline
\hline
\multicolumn{1}{c}{PSR} & \multicolumn{1}{c}{$\theta_\mathrm{max}$\tablenotemark{a}} & \multicolumn{1}{c}{$H$\tablenotemark{b}}& \multicolumn{1}{c}{$\delta$\tablenotemark{c}} & \multicolumn{1}{c}{FWHM\tablenotemark{d}} \\
\multicolumn{1}{c}{(J2000)} & (\degrees{})\\
\hline
J0631+1036   & 1.0 & $9\times10^{-7}$ & $0.44\pm0.02\pm0.0002$ &  $0.25$\\
J0659+1414   & 1.0 & $<4\times10^{-8}$ & $0.21\pm0.01\pm0.002$  &  $0.20$\\
J0742--2822  & 1.0 & $9\times10^{-6}$ & $0.61\pm0.01\pm0.001$   &  $0.10$\\
J1420--6048\tablenotemark{e}  & 0.5 & $<4\times10^{-8}$ & $0.36\pm0.02\pm0.01$   &  $0.35$\\
J1509--5850\tablenotemark{e}  & 0.5 & $<4\times10^{-8}$ & $0.31\pm0.02\pm0.02$   &  $0.40$\\
J1718--3825  & 0.5 & $<4\times10^{-8}$ & $0.42\pm0.01\pm0.006$   &  $0.20$\\
\hline
\end{tabular}
\end{center}
\tablenotetext{a}{Maximum radius from the source to include
$\gamma$-ray photons for the analysis.}
\tablenotetext{b}{Bin-independent H-test probability that the pulsation would be caused
by noise fluctuations.}
\tablenotetext{c}{Peak position (phase lag) with respect to the peak of the radio profile (the
uncertainties are the statistical and systematic errors respectively).}
\tablenotetext{d}{Full-width-half-maximum of the peak as a fraction of the pulse period.}
\tablenotetext{e}{There is some evidence, although it is statistically
not significant given the current count statistics, that the light
curves of PSRs J1420--6048 and J1509--5850 consist of two overlapping
components. If this interpretation is correct, then $\delta$ is $0.26$ and $0.44$ for the two peaks of PSR J1420--6048 and $0.18$ and $0.39$ for PSR J1509--5850.}
\end{table*}

\subsection{Temporal analysis}

The $\gamma$-ray data are collected by the Large Area Telescope (LAT;
\citealt{aaa+09}), a pair-production telescope on board {\em
Fermi}. With a large effective area ($\sim8000$ cm$^2$ above $1$ GeV,
on axis), a broad field of view ($\sim2.4$ steradian) and a high
angular resolution, this telescope is far superior to the Energetic
Gamma Ray Experiment Telescope (EGRET) on the {\em Compton Gamma Ray
Observatory} ({\em CGRO}).
The $\gamma$-ray events are time
stamped using a GPS clock on board the satellite. These arrival
times are transformed to a barycentric arrival time using the {\em
Fermi} LAT {\em Science
Tools}\footnote{http:$//$fermi.gsfc.nasa.gov/ssc/data/analysis/scitools/overview.html}
by taking into account the orbit of the satellite and the solar system
ephemerides (Jet Propulsion Laboratory DE405;
\citealt{sta98b,ehm06}). Tests have shown that the resulting precision
is accurate to at least a few microseconds \citep{sgc+08}.

In this paper we use photons which were collected between 2008 June 30
and 2009 May 22. Only so-called ``diffuse'' class events were used
\citep{aaa+09}, which are those with the highest probability to be
caused by $\gamma$-rays from the source. All $\gamma$-rays events with
a reconstructed zenith angle larger than $\degrees{105}$ were ignored to avoid
the intense $\gamma$-ray background caused by cosmic-ray
interactions in the Earth's atmosphere.

To produce the $\gamma$-ray light curves we used photons within
an energy-dependent radius $\theta\le0.8\times
E_{\mathrm{GeV}}^{-0.75}$~degrees of the pulsar position, requiring a
radius of at least 0\fdg35, but not larger than the
$\theta_\mathrm{max}$ of 0\fdg5 or 1\fdg0 as shown in Table
\ref{tableGammaLightcurves}. See the LAT pulsar catalog paper
\citep{aaa+09e} for a detailed discussion of the selection criteria.
This selection maximizes the signal-to-noise ratio over the broad
energy range covered by the LAT.  In all cases the background is
estimated from off-pulse bins from a \degrees{1}--\degrees{2} ring
around the pulsar using the same energy-dependent cut.

\begin{table}[t]
\caption{\label{RefTableRadio}Radio timing parameters. 
}
\begin{center}
\begin{tabular}{llrr@{$\pm$}l}
\hline
\multicolumn{1}{c}{PSR} & Obs\tablenotemark{a} & rms & \multicolumn{2}{c}{DM} \\
\multicolumn{1}{c}{(J2000)} & & ($\mu s$) & \multicolumn{2}{c}{(cm$^{-3}$ pc)}\\
\hline
\hline
J0631+1036  & J/N         & 51   & $125.36$&$0.01$  \\
J0659+1414  & P/N         & 427  & $13.7$&$0.2$     \\
J0742--2822 & P/J/N       & 210  & $73.790$&$0.003$ \\
J1420--6048 & P           & 516  & $358.8$&$0.2$    \\
J1509--5850 & P           & 1068 & $140.6$&$0.8$    \\
J1718--3825 & P           & 434  & $247.88$&$0.09$  \\
\hline
\end{tabular}
\end{center}
\tablenotetext{a}{The radio observatories (P = Parkes, J = Jodrell
Bank\\and N = Nan\c{c}ay) involved in the radio timing.}
\end{table}

The pulsars discussed in this paper are regularly observed in the
radio band near 1.4 GHz by the {\em Fermi} Timing Consortium (see
Table \ref{RefTableRadio}). The observatories involved are the Parkes
64-m radio telescope in Australia, the Lovell 76-m telescope at the
Jodrell Bank observatory near Manchester in England, and the 94-m
(equivalent) Nan\c{c}ay radio telescope near Orleans, France. The
timing program at Parkes is described in \cite{wjm09} and typically
involves monthly observations.  Nan\c{c}ay \citep{ctd+09} and Jodrell
Bank \citep{hlk+04} observations are made, on average, every 5 to 9
days.

The times of arrival (TOAs) for PSRs J1420--6048, J1509--5850 and
J1718--3825 were obtained solely from Parkes data.
For the other pulsars the TOAs obtained from the different telescopes
were combined before making an ephemeris for the spin behavior of the
neutron star. The TOAs were compared with an initial timing solution
using {\sc TEMPO2} \citep{hem06} producing timing residuals which then
were fit for the spin-frequency and its time derivative, as well as
for instrumental offsets between data from different
observatories. Most of the pulsars showed strong additional deviations
in their spin behavior, known as timing noise (e.g. \citealt{hlk+04}). This timing noise is
modelled by either adding higher-order spin-frequency derivatives or
by using the fitwaves algorithm within {\sc TEMPO2}.

The resulting timing model allows an accurate assignment of a
rotational phase to the $\gamma$-ray photons (using {\sc
TEMPO2}), thereby constructing the light curves. The timing parameters
used in this work will be made available on the servers of the
\textit{Fermi} Science Support
Center\footnote{http:$//$fermi.gsfc.nasa.gov/ssc/data/access/lat/ephems/}.

An important parameter necessary to align the radio profile with the
$\gamma$-ray light curve is the dispersion-measure (DM), which
quantifies the frequency-dependent delay of the radio emission caused
by the interstellar medium. The DM of PSR J0631+1036 was
measured by comparing the TOAs at two widely separated frequencies
using Jodrell Bank data (0.6 GHz and 1.4 GHz) after the templates used
were carefully aligned so that the components at the two frequencies
coincide in pulse phase. The DM of the other five pulsars were obtained
by measuring the delay across the 256 MHz band using Parkes data (see
\citealt{wjm09} and Table \ref{RefTableRadio}). The systematic error
in the alignment of the radio and the $\gamma$-ray light curves is the
combination of the DM uncertainty and the rms scatter of the radio timing
residuals.

The shapes of the light curves (consisting of all photons above 100
MeV) were fitted using Gaussian functions resulting in peak positions
and the error bars on the light curves were taken to be the
square-root of the number of photons in each phase bin.  The results
of the fitting of the light curves together with the
full-width-half-maxima (FWHM) estimated by eye are summarized in Table
\ref{tableGammaLightcurves}. This table includes the bin-independent
H-test probability \citep{drs89}, which measures the probability that
the observed $\gamma$-ray light curves are caused by noise
fluctuations.

\subsection{Spectral analysis}

 The spectral analysis uses the same first 6 months of
\textit{Fermi} data as the LAT pulsar catalog \citep{aaa+09e}. The LAT
``gtlike'' science
tool\footnote{\label{footnoteFermiDocumentation}http://fermi.gsfc.nasa.gov/ssc/data/analysis/documentation/}
performs a maximum likelihood analysis \citep{mbc+96} to fit
phase-averaged spectra for the six pulsars.  Instrument Response
Functions (IRFs) allow proper treatment of the direction and energy of
each event. We used ``Pass 6 v3'', a post-launch IRF update that
addresses inefficiencies correlated with the trigger
rate\footnotemark[3].  Angular
resolution is poor at low energies: at 100 MeV and normal detector
incidence, 68\% of the photons from a point source have reconstructed
directions within $\sim 5^\circ$ of the true direction, decreasing to
$\sim 0.2^\circ$ at 10 GeV.  Therefore, the likelihood analysis must
model not just the pulsar under study, but all neighboring $\gamma$-ray
sources as well.  We applied the analysis used for the 3-month
\textit{Fermi} Bright Source List (BSL; \citealt{aaa+09b}) with an updated model for the Galactic diffuse emission to this 6-month
data set.  Then, as for the BSL, we extract events in a circle of
radius 10$\degr$ around each pulsar.  The likelihood model includes
all sources up to 17$\degr$ from each pulsar, with spectral parameters
fixed to the values obtained from the BSL analysis for those more than
3$^{\circ}$ away.  Spectral parameters for the pulsar, as well as
for sources within 3$^{\circ}$, are left free in the fit.  Galactic
diffuse emission was modeled using a GALPROP \citep{smr04} calculation
designated {\texttt{54\_77Xvarh7S}}, very similar to that available
from the $Fermi$ Science Support
Center\footnotemark[3].

Bright $\gamma$-ray pulsars like Vela \citep{aaa+09d} or the Crab
\citep{aaa+09g} are observed to have spectra which are well described
by exponentially cutoff power-law models of the form
\begin{equation}
\frac{{\rm d} N}{{\rm d} E} = K E_{\rm GeV}^{-\Gamma}
                              \exp \left(- \frac{E}{E_{\rm cutoff}} \right)
\label{expcutoff}
\end{equation}
in which the three parameters are the normalization $K$ of the
differential flux (in units of ph cm$^{-2}$ s$^{-1}$ MeV$^{-1}$), the
spectral index, $\Gamma$, and the cutoff energy, $E_{\rm cutoff}$. The
energy at which the differential flux is defined is arbitrary. We
choose 1 GeV because it is, for most pulsars, close to the energy at
which the relative uncertainty on the differential flux is
minimal. 
The spectra were fitted separately using a power-law and a
power-law plus exponential cutoff. The difference $\Delta$ of the
$\log(\mathrm{likelihood})$ for the two fits determines the
significance $\sigma_\mathrm{cutoff}$ for the existence of an energy
cutoff, which is defined to be the test statistic difference
$\sqrt{2\Delta}$. For pulsars with a $\sigma_\mathrm{cutoff}<3$ the
power-law with a cutoff did not result in a significantly better fit
compared to a simple power-law, hence the cutoff energy is
unconstrained.

\begin{table*}[t]
\caption{\label{tableGammaSpec}The parameters of the spectral analysis
of the $\gamma$-ray data\tablenotemark{a}. 
}
\begin{center}
\begin{tabular}{lr@{$\pm$}lccc}
\hline
\hline
\multicolumn{1}{c}{PSR} & \multicolumn{2}{c}{Energy Flux ($E > 100$ MeV)} & \multicolumn{1}{c}{$\Gamma$} & \multicolumn{1}{c}{$E_{\rm cutoff}$} & \multicolumn{1}{c}{$\sigma_\mathrm{cutoff}$} \\
\multicolumn{1}{c}{(J2000)}& \multicolumn{2}{c}{($10^{-11}$ erg\,cm$^{-2}$\,s$^{-1}$)} & & \multicolumn{1}{c}{(GeV)} & \\
\hline
J0631+1036   & \hspace*{14mm}$3.04$&$0.51$ & $1.38\pm0.35$ & $3.6\pm1.8$ & 3.2 \\	
J0659+1414   & $3.17$&$0.31$ & $2.37\pm0.42$ & $0.7\pm0.5$ & 2.6 \\	
J0742--2822  & $1.83$&$0.36$ & $1.76\pm0.40$ & $2.0\pm1.4$ & 2.0 \\
J1420--6048  & $15.9$&$2.8$  & $1.73\pm0.20$ & $2.7\pm1.0$ & 4.6 \\
J1509--5850  & $9.7 $&$1.0$  & $1.36\pm0.23$ & $3.5\pm1.1$ & 5.1 \\
J1718--3825  & $6.8 $&$1.7$  & $1.26\pm0.62$ & $1.3\pm0.6$ & 4.4 \\
\hline
\end{tabular}
\end{center}
\tablenotetext{a}{The spectral analysis is based on the first 6 months of {\em Fermi}
data as presented in the LAT pulsar catalog paper \citep{aaa+09e}.}
\end{table*}

\begin{table*}[t]
\caption{\label{tableGammaLum}The $\gamma$-ray luminosity and the
efficiency, which are functions of the flux correction factor $f_\Omega$ and the pulsar distance $D$.
}
\begin{center}
\begin{tabular}{lr@{$f_\Omega$}lr@{$f_\Omega$}ll}
\hline
\hline
\multicolumn{1}{c}{PSR} & \multicolumn{2}{c}{$L_\gamma$} & \multicolumn{2}{c}{$\eta$} \\
\multicolumn{1}{c}{(J2000)} & \multicolumn{2}{c}{($10^{35}$ erg\,s$^{-1}$)} & \multicolumn{2}{c}{}\\
\hline
J0631+1036   & $(0.036\pm0.006)$&$\left(D/1\;\mathrm{kpc}\right)^2$        & $(0.021\pm0.004)$&$\left(D/1\;\mathrm{kpc}\right)^2$\\	
J0659+1414   & $(0.0032\pm0.0003)$&$\left(D/0.288 \;\mathrm{kpc}\right)^2$ & $(0.0084\pm0.0008)$&$\left(D/0.288 \;\mathrm{kpc}\right)^2$\\	
J0742--2822  & $(0.09\pm0.02)$&$\left(D/2 \;\mathrm{kpc}\right)^2$         & $(0.06\pm0.01)$&$\left(D/2 \;\mathrm{kpc}\right)^2$\\
J1420--6048  & $(6\pm1)$&$\left(D/5.6 \;\mathrm{kpc}\right)^2$             & $(0.06\pm0.01)$&$\left(D/5.6 \;\mathrm{kpc}\right)^2$ \\
J1509--5850  & $(0.75\pm0.08)$&$\left(D/2.5 \;\mathrm{kpc}\right)^2$       & $(0.15\pm0.02)$&$\left(D/2.5 \;\mathrm{kpc}\right)^2$\\
J1718--3825  & $(1.1\pm0.3)$&$\left(D/3.6 \;\mathrm{kpc}\right)^2$         & $(0.09\pm0.03)$&$\left(D/3.6 \;\mathrm{kpc}\right)^2$ \\
\hline
\end{tabular}
\end{center}
\end{table*}

The observed $\gamma$-ray energy flux $F_\mathrm{obs}$ is the 
integral above 100 MeV of the fitted spectral shape times the energy.
The luminosity is then
\begin{equation}
L_\gamma=4\pi f_\Omega F_\mathrm{obs}D^2,
\end{equation}
where $D$ is the distance and $f_\Omega$ is the flux correction factor
which depends on the beaming fraction (e.g. \citealt{wrwj09}).  For
outer magnetospheric models $f_\Omega$ is thought to be $\sim1$, which is the value we assume throughout this paper except
in Table \ref{tableGammaLum}.
This luminosity can then be compared to the
spin-down energy loss rate of the pulsar to obtain the $\gamma$-ray
efficiency
\begin{equation}
\eta=L_\gamma/\dot{E}.
\end{equation}
The results of the spectral analysis are summarized in Table
\ref{tableGammaSpec} and the derived luminosities and efficiencies can
be found in Table \ref{tableGammaLum}. Note that the distances
to all the pulsars except PSR J0659+1414 are highly uncertain, as
described in more detail below.

 Uncertainties on the effective area ($\le$ 5\% near 1 GeV, 10\%
below 0.1 GeV and 20\% over 10 GeV) and uncertainties in the Galactic
diffuse emission model dominate the systematic uncertainties on the
spectral results as described in the LAT pulsar catalog
\citep{aaa+09e}.
The spectral parameter uncertainties are $\delta\Gamma =
(+0.3,\,-0.1)$, $\delta E_{\rm cutoff} = (+20\%,\,-10\%)$, and $\delta
F_\mathrm{obs} = (+20\%,\,-10\%)$.  The bias on the integral energy flux is
somewhat less than that of the integral photon flux, due to the
weighting by photons in the energy range where the effective area
uncertainties are smallest. We do not sum these uncertainties in
quadrature with the others, since a change in instrument response will
tend to shift all spectral parameters similarly.  

\section{Deriving emission geometries from radio data}
\label{SctMethods}

Two important angles used to describe the emission geometry of
pulsars are the angle $\alpha$ between the magnetic axis and the
rotation axis and the angle $\zeta$ between the line-of-sight and the
rotation axis. A related angle is the impact parameter $\beta =
\zeta-\alpha$, which is the angle between the line-of-sight and the
magnetic axis at its closest approach. These angles can be inferred by
applying the rotating vector model (RVM; \citealt{rc69a}) to the
position angle (PA) of the linear polarization observed in the radio
band. This model predicts the PA of the linear polarization $\psi$ to
depend on the pulse phase $\phi$ as
\begin{eqnarray}
\label{EqRVM}
\nonumber \tan\left(\psi-\psi_0\right)=\hspace*{44mm}\\\frac{\sin\alpha\;\sin\left(\phi-\phi_0\right)}{\sin\zeta\;\cos\alpha-\cos\zeta\;\sin\alpha\;\cos\left(\phi-\phi_0\right)},
\end{eqnarray}
where $\psi_0$ and $\phi_0$ are the PA and pulse phase corresponding
to the intersection of the line-of-sight with the fiducial plane (the
plane containing the rotation and magnetic axis) if the emission
height $h_\mathrm{em}$ is small compared to the light cylinder
distance. In this model the PA-swing is an S-shaped curve and its
inflection point occurs at $\phi_0$.

It is found that the degree of linear polarization is correlated with
$\dot{E}$ such that virtually all pulsars with
$\dot{E}>2\times10^{35}$ erg\,s$^{-1}$ have a linear polarization
fraction over 50\% (\citealt{wj08b}; see also
\citealt{qmlg95,hlk98,cmk01,jw06}). Moreover most of these pulsars
have smooth PA-swings, making it relatively easy to apply the RVM
model to the pulsars presented in this
paper.

If the emission profile is symmetric around the magnetic axis, then
the inflection point coincides with the middle of the pulse
profile. However, co-rotation of the emitting region causes the
inflection point to be delayed with respect to the pulse profile. This
pulse phase difference $\Delta\phi$ between the middle of the profile
and the inflection point of the PA-swing can be used to estimate the
emission height (\citealt{bcw91})
\begin{equation}
\label{EqBCW}
h_\mathrm{PA} = \frac{P\,c}{8\pi }\Delta\phi = \frac{1}{4}R_\mathrm{LC}\Delta\phi,
\end{equation}
where $P$ is the spin period of the pulsar, $c$ is the speed of light
and $R_\mathrm{LC}$ is the light cylinder radius.  Because the
relative shift of the PA-swing with respect to the profile is
independent of $\alpha$ and $\zeta$ \citep{drh04}, Equation
\ref{EqRVM} can be used to fit the PA-swing even for moderate emission
heights \citep{dyk08}. Use of these equations ignores effects of
rotational sweepback of the magnetic field lines \citep{dh04},
propagation effects in the pulsar magnetosphere
(e.g. \citealt{pet06}), current-induced distortions of the magnetic
field and the effects of a finite emission height spread and emission
height differences \citep{dyk08}.

Both the PA-swing and the observed pulse width contain information
about the geometry of the pulsar. Under the assumption that the radio
beam is symmetric about the magnetic axis, the pulse width $W$ (which
we take to be the full phase range over which we see emission) is
related to the half opening angle $\rho$ of the beam via
\begin{eqnarray}
\label{EqCosRho}
\cos\rho = \cos\alpha \cos\zeta+\sin\alpha \sin\zeta \cos\left(W/2\right),
\end{eqnarray}
(\citealt{ggr84,lk05}). Therefore, if we know the value of $\rho$ for a given
pulsar, Equation \ref{EqCosRho} can be used as an additional
constraint on Equation \ref{EqRVM} to narrow down the allowed region
in $\alpha-\beta$ space. The value of $\rho$ can be estimated from the
emission height via
\begin{equation}
\label{EqRho}
\rho=\sqrt{\frac{9\pi h_\mathrm{em}}{2Pc}},
\end{equation}
assuming a radio beam centred about the magnetic axis which is
enclosed by the last open dipole field lines (e.g. \citealt{lk05}). In
general, although there is some evidence for conal rings centered on
the magnetic axis \citep{ran83}, many pulsars are patchy
(e.g. \citealt{lm88,hm01,kjw+09,ww09}), making both $\rho$ and
$h_\mathrm{PA}$ uncertain.  This uncertainty is likely to dominate
the total uncertainty of the radio analysis and its impact is
discussed for the individual pulsars in the next section. Because of
poorly understood systematics it is impossible to come up with 
sensible error bars on the derived values describing the geometry.

In summary, first we can use the RVM to fit the PA-swing (Equation
\ref{EqRVM}), resulting in contours defining the allowed
$\alpha-\beta$ parameter space. Secondly, from the offset of the
PA-swing with respect to the total intensity profile, we can estimate
an emission height (Equation \ref{EqBCW}), which can be translated to
an opening angle of the radio beam (Equation \ref{EqRho}, assuming
$h_\mathrm{em}=h_\mathrm{PA}$). This opening angle corresponds to
another contour in $\alpha-\beta$ parameter space (Equation
\ref{EqCosRho}), which in the ideal case would match with the RVM
contours. If not, then at least one of the assumptions must be
incorrect, most likely indicating that the radio beam is asymmetric
with respect to the magnetic axis.

Additional constraints on the emission geometry can be obtained from
measurements of the termination shock of the surrounding pulsar wind
nebula which can provide a relatively model-independent estimate of
the viewing angle $\zeta$ \citep{nr08}; unfortunately such
measurements are not yet available for the pulsars discussed here.

\section{Results on the individual pulsars}
\label{SctResults}

\subsection{PSR J0631+1036}

\subsubsection{The pulsar and its surroundings}

\object{PSR J0631+1036} was discovered as a young pulsar by
\cite{zcwl96} in a radio search targeting {\em Einstein} IPC X-ray
sources. Its DM is very high for a pulsar in the Galactic anticenter
and this is argued to be caused by the foreground star-forming region
3 Mon \citep{zcwl96}. In addition this pulsar could be interacting
with (or be embedded in) dark cloud LDN 1605 and therefore the
distance derived from the DM ($3.6\pm1.3$ kpc according to the
\citealt{cl02} model) is possibly overestimated. Following
\cite{zcwl96} we adopt a distance of 1 kpc consistent with the
observed X-ray absorption.  No pulsar wind nebula (PWN) has been found
for this pulsar at radio wavelengths \citep{gsf+00}.

\begin{figure}[t!]
\begin{center}
\includegraphics[angle=0,width=\figwidth, trim=30 0 50 0, clip=true]{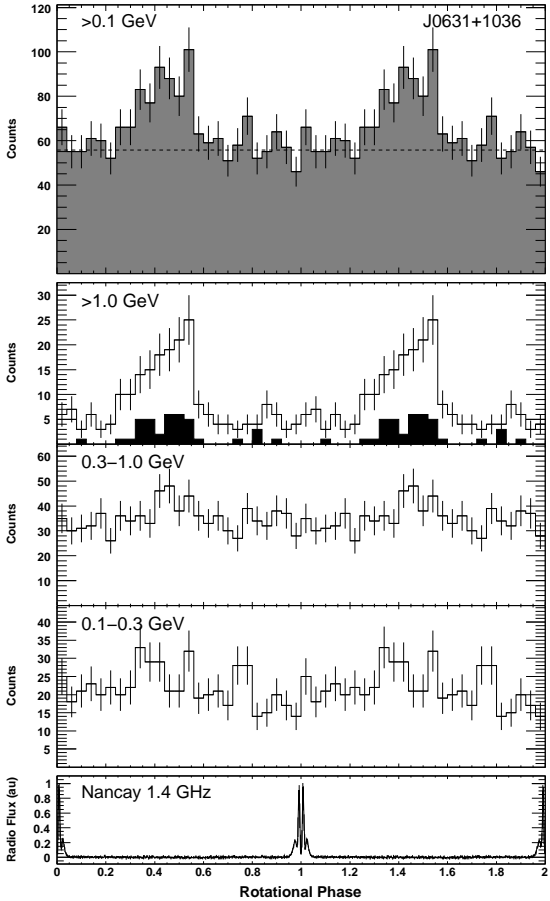}
\end{center}
\caption{\label{Fig0631}The {\em Fermi} $\gamma$-ray light curves of
PSR J0631+1036 in different energy bands showing two full rotational
periods with 25 bins per period. The photons above 3 GeV are shown in
black in the second panel from the top. The background estimate is
shown as a dashed line in the top panel. The bottom panel shows the
phase-aligned radio profile. }
\end{figure}

\begin{figure}[t]
\epsscale{.70}
\begin{center}
\includegraphics[angle=270,width=\figwidth]{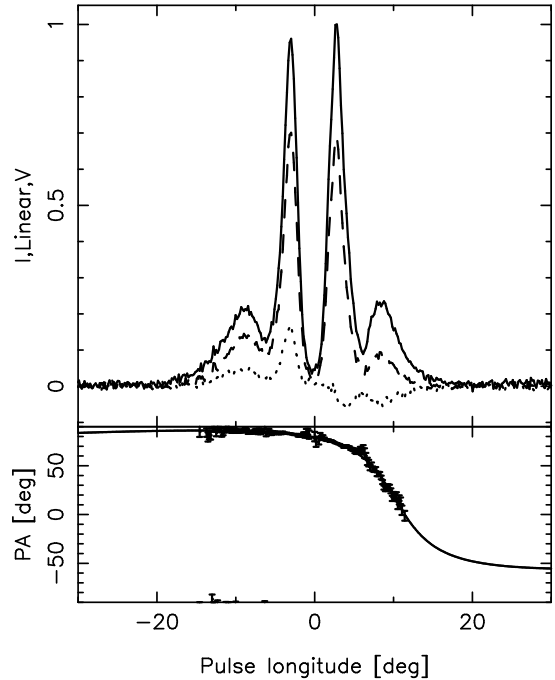}
\end{center}
\caption{\label{FigProfile0631}The
pulse profile at 1398 MHz observed at Nan\c{c}ay (black) of PSR
J0631+1036, as well as the degree of linear polarization (dashed) and
circular polarization (dotted). The bottom panel shows the PA of the
linear polarization (if detected above $3\sigma$) and an RVM fit.}
\end{figure}

\subsubsection{$\gamma$-rays}

PSR J0631+1036 clearly shows $\gamma$-ray pulsations (see Figure
\ref{Fig0631}) and the light curve features a single broad peak (FWHM
$0.25$ in rotational phase) which lags the radio profile by
$0.44$ (see Table \ref{tableGammaLightcurves}). 
A power law in combination with an exponential cutoff fits the
$\gamma$-ray spectrum significantly better than a single power law,
although the cutoff energy cannot be accurately determined.

\cite{zcwl96} claimed a $\gamma$-ray detection by EGRET of PSR
J0631+1036. The detection was very marginal and their light curve does
not really resemble that seen by {\em Fermi}. Their estimated
$\gamma$-ray flux 
is an order of magnitude larger than that obtained from the {\em
Fermi} data (see Table \ref{tableGammaSpec}).

\subsubsection{X-rays}

At the pulsar position a faint {\em ROSAT} PSPC X-ray source has been
found, which was too weak to search for pulsations
\citep{zcwl96}. Sinusoidal X-ray pulsations were claimed by
\cite{tsn+01} using {\em ASCA} data. However, an {\em XMM-Newton}
observation appears to show that the X-ray point source is not
associated with the pulsar and the pulsations were not confirmed
\citep{kcc+02}. The derived upper limit for a X-ray point source
is $1.1\times10^{30}$ erg\,s$^{-1}$ (0.5-2.0 keV) for the assumed distance
of 1 kpc. This luminosity is low compared to the \cite{bt97}
relationship between the X-ray luminosity and $\dot{E}$, which can be
seen as evidence for a larger distance to the pulsar. 

\subsubsection{Radio}

The pulsar's radio spectrum is relatively flat from 1.4 up to at least
6.2 GHz (unpublished Parkes
data\footnote{\label{RefSimonwww}www.atnf.csiro.au/people/joh414/ppdata})
and its radio profile is highly polarized \citep{zcwl96,wj08b}.
The radio profile consists of four components (see Figure
\ref{FigProfile0631}) and the outer components are strongest at low
frequencies. The radio profile shows a remarkable deep minimum at the
pulse phase of the symmetry point. Despite the complex structure of
the profile, it is highly mirror symmetric, not only in its shape, but
also in the spectral indices of the different components.

\begin{figure}[t]
\epsscale{.70}
\begin{center}
\includegraphics[angle=270,width=\figwidth]{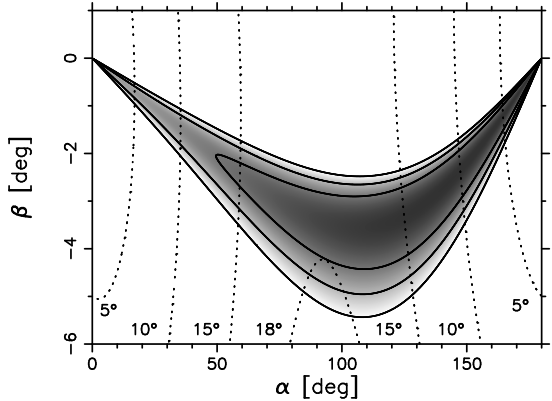}
\end{center}
\caption{\label{FigRVM0631}The $\chi^2$ map of the RVM fit of
J0631+1036 (in gray scale, a darker color indicates a lower
$\chi^2$). The lowest reduced $\chi^2$ is 1.45 and the solid contours
correspond to reduced $\chi^2$ values that are two, three and four
times larger. The half opening angles $\rho$ of the radio beam derived from
the observed pulse width are overplotted (dotted contours).}
\end{figure}

We fitted the RVM model (Equation \ref{EqRVM}) to the radio
polarization data. The resulting $\chi^2$ map is shown in gray scale
and in the black contours in Figure \ref{FigRVM0631}. One can see
a ``banana-shaped'' region in $\alpha-\beta$ space which contains
valid solutions for the PA-swing, showing that $\alpha$ is not
constrained.

The remarkable degree of mirror symmetry of the radio profile can be
seen as evidence that the radio beam itself has a high degree of
symmetry and is centered at the magnetic axis. Under this assumption
the radio emission height follows from the offset of the steepest
gradient of the PA swing and the mirror point of the profile (Equation
\ref{EqBCW}) and is found to be $h_\mathrm{PA}=600$ km, which is
slightly lower than found by \cite{wj08b} using data with a lower
signal-to-noise ratio.
Using this emission height in Equation \ref{EqRho} we expect a half
opening angle of the beam $\rho$ of \degrees{18} if the emission comes
from the last open field line.
In Figure \ref{FigRVM0631} the contours of $\rho$ derived from the
observed pulse width are overplotted (dotted). One can see that such a
large value of $\rho$ suggests that $\alpha$ is close to \degrees{90}
and $\beta\sim\degrees{-4}$. If only a fraction
of the open field lines produces radio emission (i.e. if the beam
is patchy) then $\alpha$ could be smaller.

\subsection{PSR J0659+1414 (B0656+14)}

\subsubsection{The pulsar and its surroundings}

\object{PSR J0659+1414} was discovered in the second Molonglo pulsar
survey as a radio pulsar \citep{mlt+78} and this pulsar is the slowest
rotating and has the lowest $\dot{E}$ of the pulsars discussed in this
paper. The distance of the pulsar is well known via parallax
measurements using very long baseline interferometry
($288^{+33}_{-27}$ pc; \citealt{btgg03}). PSR J0659+1414 is associated
with the Monogem ring \citep{tbb+03}, a bright diffuse \degrees{25}
diameter supernova remnant easily visible in soft X-ray images of the
sky. It has a possible pulsar wind nebula in optical \citep{szk+06}
and X-rays \citep{ms02}.

\subsubsection{$\gamma$-rays}

\begin{figure}[t!]
\begin{center}
\includegraphics[angle=0,width=\figwidth, trim=30 0 50 0, clip=true]{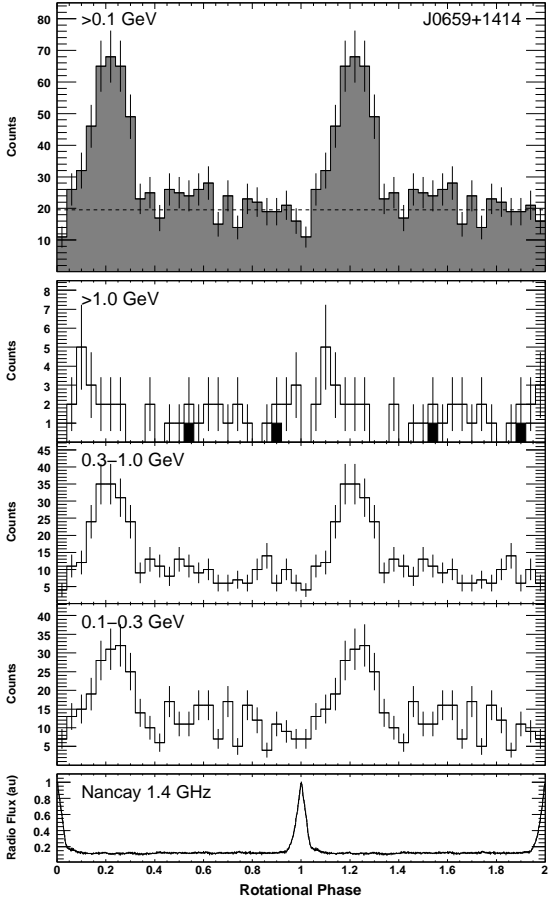}
\end{center}
\caption{\label{Fig0656}As in Figure \ref{Fig0631}, but for PSR J0659+1414.
}
\end{figure}

There was a marginal detection of pulsed $\gamma$-rays from this pulsar
by EGRET \citep{rfk+96}. The light curve observed by {\em Fermi} (see
Figure \ref{Fig0656}) is similar to that seen by EGRET, but the
$\gamma$-ray background is much lower in the {\em Fermi} data because
of its superior angular resolution. The light curve is single peaked
(FWHM $0.20$ in rotational phase) and lags the radio peak by
$0.21$ in phase. 

The pulsar is very weak above 1 GeV, showing that its spectrum is
extremely soft. Indeed this pulsar appears to have the steepest
spectrum and the lowest cutoff energy of the pulsars in this
paper (see Table \ref{tableGammaSpec}). Although the cutoff
appears to be within the energy range of the LAT detector, a power law
plus exponential cutoff does not describe the data better than a
single power law. The $\gamma$-ray efficiency of PSR J0659+1414 is
very low (see Table \ref{tableGammaLum}). Because the distance
to the pulsar is well determined, the low efficiency cannot be caused
by an incorrectly estimated luminosity.

\subsubsection{X-rays}

\begin{figure}[t!]
\begin{center}
\includegraphics[angle=270,width=\figwidth]{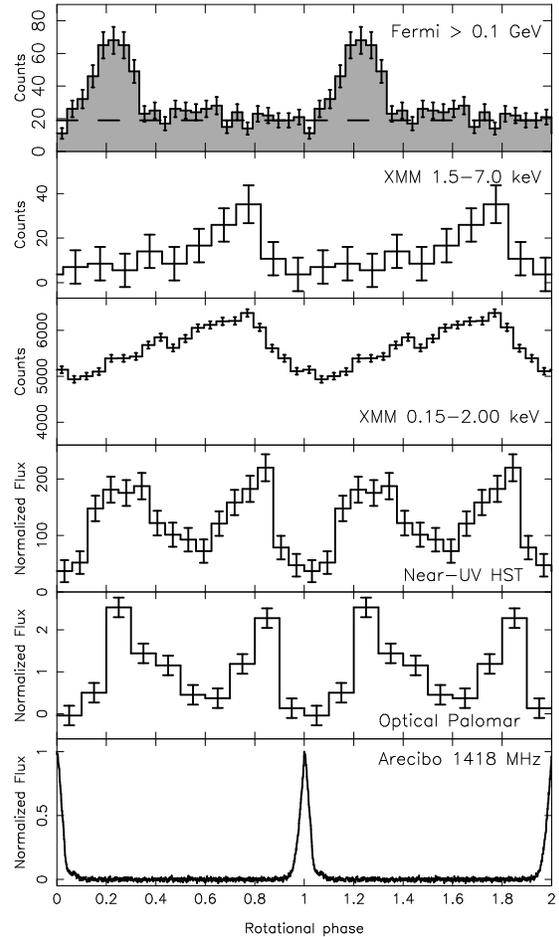}
\end{center}
\caption{\label{Fig0656Compilation}The phase-aligned light curves of
PSR J0659+1414 at multiple wavelengths. The {\em Fermi} light curve
(with its background level indicated by the dashed line) is compared
with the XMM observations \citep{dcm+05}, shown in two energy ranges
encompassing thermal (0.15-2 keV) and non-thermal (1.5-7 keV)
components (note that for the lower panel the vertical axis does not
start at zero, indicating a large unpulsed fraction of the
emission). The near-UV light curve is from \cite{ssl+05}, who
aligned the light curve by making use of the similarity with the
optical light curve. The
background-subtracted optical light curve is from \cite{kmmh03} and
the radio profile is from \cite{ew01}. 
}
\end{figure}

PSR J0659+1414 is one of the brightest isolated neutron stars in the
X-ray sky \citep{cmhm89} and it is one of the ``Three Musketeers''
(the others being Geminga and PSR B1055--52; \citealt{bt97}). The
X-ray emission is a combination of thermal (blackbody) and non-thermal
(power law) emission (e.g. \citealt{dcm+05}) and is consistent with a
cooling middle-aged neutron star (e.g. \citealt{bt97}). At soft X-rays
the pulse fraction is low and the pulsations are sinusoidal (see Figure
\ref{Fig0656Compilation}), typical for thermal emission from the
surface of a neutron star with a non-uniform temperature distribution
(i.e. hotter polar caps).
At higher energies ($>1.5$ keV), where the non-thermal
component dominates, the pulsed fraction increases and the profile
becomes single peaked.

A relatively aligned geometry is consistent with the sinusoidal
soft X-ray profile, which suggests only one pole is visible from Earth
and the low amount of modulation in soft X-rays also hints toward an
aligned rotator \citep{dcm+05}. However, there are some complications
in modelling of the soft X-rays, because there is an apparent
anti-correlation between the hot and cool blackbody component of the
thermal part of the emission that is not easily understood without
invoking significant multipole components of the magnetic field or
magnetospheric reprocessing of thermal photons. In addition the
best-fitting emitting radius of $\sim$21 km (using the very accurate
radio VLBI parallax measurement) is unlikely given the expectations
for a standard neutron star.

\subsubsection{Optical}

PSR J0659+1414 is seen at optical wavelengths
\citep{cbm94}, allowing the study of the
(non-thermal) optical pulsations (e.g. \citealt{kmmh03}).
The optical light curve is double peaked (see Figure
\ref{Fig0656Compilation}) and is very similar to that seen in near-UV
\citep{ssl+05}. The first optical peak following the radio pulse is
aligned with the single $\gamma$-ray peak, while the second optical
peak following the radio peak is aligned with the peak seen in
(non-thermal) X-rays above 1.5 keV.

\subsubsection{Radio}

\begin{figure}[t]
\epsscale{.70}
\begin{center}
\includegraphics[angle=270,width=\figwidth]{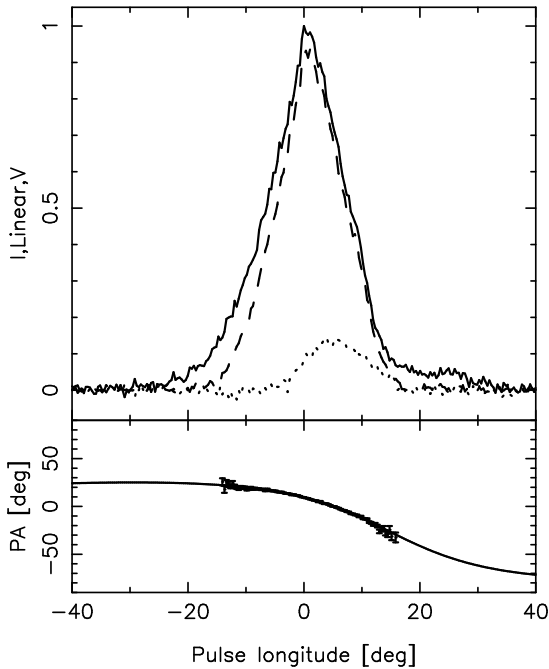}
\end{center}
\caption{\label{FigProfile0656}As in Figure \ref{FigProfile0631}, but
for PSR J0659+1414 at 1418 MHz.
These Arecibo data are taken from
\cite{wcl+99}.}
\end{figure}

\begin{figure}[t]
\epsscale{.70}
\begin{center}
\includegraphics[angle=270,width=\figwidth]{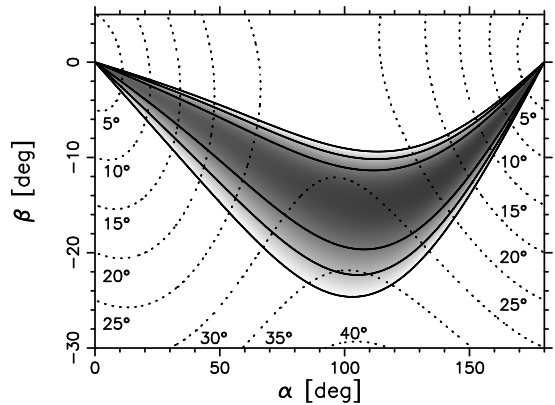}
\end{center}
\caption{\label{FigRVM0656}As in Figure \ref{FigRVM0631}, but for PSR
J0659+1414. The lowest reduced $\chi^2$ is 0.93.
}
\end{figure}

The radio profile of PSR J0659+1414 is roughly triangular at 1.4 GHz
(see Figure \ref{FigProfile0656}) with a weak shoulder at the trailing
edge. The weak shoulder is associated with weak and broad radio
pulses, while the radio pulses in the triangle have a ``spiky''
appearance \citep{wws+06}. The strongest of these radio pulses are
argued \citep{wsr+06} to be similar to transient radio bursts seen
from the so-called rotating radio transients (RRATs;
\citealt{mll+06}). The radio emission is almost completely linearly
polarized at 1.4 GHz, but there is significant depolarization at the
leading edge. There is also significant circular polarization, which
is strongest at the trailing edge of the profile. Curiously, at 6.2
and 8.4 GHz the profile is completely depolarized (\citealt{jkw06} and
unpublished Parkes data\footnotemark[4]).

Figure \ref{FigRVM0656} shows the $\chi^2$ map of the RVM fit for a
high signal-to-noise radio profile taken from \cite{wcl+99}. The
results are consistent with previous results
(e.g. \citealt{lm88,ran93b,ew01}) and it is immediately clear that, as
is often the case, $\alpha$ is unconstrained.

The steepest gradient of the PA-swing lags the peak of the radio peak
by $14\fdg9\pm0\fdg7$ \citep{ew01}, suggesting an emission height of
1200 km (Equation \ref{EqBCW}) and $\rho=\degrees{22}$ (using Equation
\ref{EqRho}), implying that $\alpha$ should be $\sim\degrees{50}$ (see
Figure \ref{FigRVM0656}). The weak shoulder at pulse phases
$\degrees{18}$--$\,\degrees{30}$ in Figure \ref{FigProfile0656} could
indicate that the peak of the profile leads the pulse phase
corresponding to fiducial plane, which would suggest that
$\alpha\lesssim\degrees{50}$.
A relatively aligned geometry would be in line with the above
discussion about the thermal X-rays. The optical light curve is
linearly polarized and optical PA points can be measured
\citep{kmmh03}. Unfortunately, while the optical linear polarization
is potentially useful, the limited data (three phase bins) and large
errors make it impossible at present to use these data to constrain
the RVM fit.
Improved optical polarization measurements, however, have the
potential to greatly refine our geometrical knowledge of this
important pulsar.

\subsection{PSR J0742--2822 (B0740--28)}

\begin{figure}[t!]
\begin{center}
\includegraphics[angle=0,width=\figwidth]{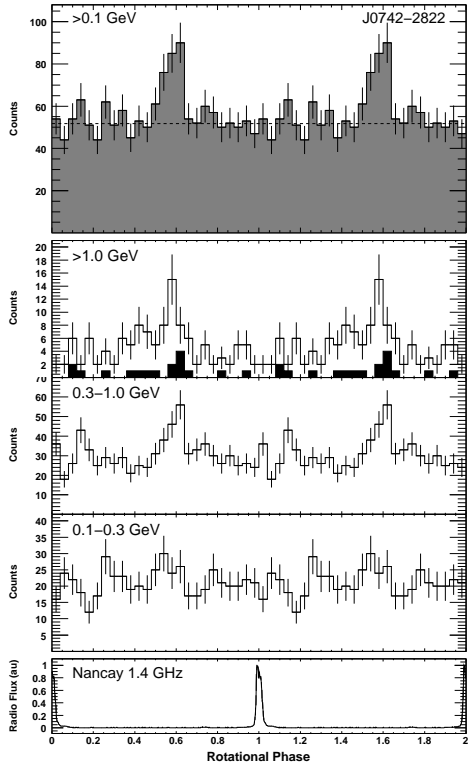}
\end{center}
\caption{\label{Fig0742}As in Figure \ref{Fig0631}, but for PSR J0742--2822.
}
\end{figure}

\subsubsection{The pulsar and its surroundings}

\object{PSR J0742--2822} was discovered as a radio pulsar by
\cite{fss73}.
\cite{kjww95} determined a kinematic distance between $2.0\pm0.6$ and
$6.9\pm0.8$ kpc, which was higher than the distance derived from the
DM according to the \citealt{tc93} model (1.9 kpc). This was argued
by \cite{kjww95} to be caused by an overestimation of the electron
density of the Gum Nebula in this position. 
There is a steep gradient in the electron density in this
direction in the \cite{cl02} model, which was constructed such
that its predicted distance is consistent with the kinematic
distance. We therefore adopt a distance of 2 kpc with the note that it
could possibly be as distant as 7 kpc.

\subsubsection{$\gamma$-rays}

PSR J0742--2822 has a relatively narrow single peak in $\gamma$-rays,
especially above 1 GeV (see Figure \ref{Fig0742}). The $\gamma$-ray
peak lags the radio peak by $0.61$ in phase.
In contrast to PSR J0659+1414 this pulsar is much weaker at low
energies and is not detected below 300 MeV. Note that if the pulsar is
at a distance of 7 kpc (the upper limit of the kinematic distance),
the $\gamma$-ray efficiency would be $\sim70$\%. This large efficiency
may indicate that the pulsar is nearer than the distance upper limit.

\subsubsection{Radio}

\begin{figure}[t]
\begin{center}
\includegraphics[angle=270,width=\figwidth]{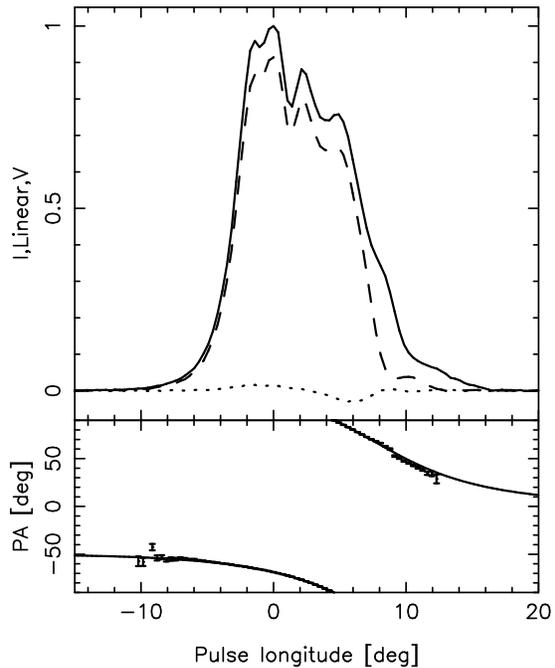}
\end{center}
\caption{\label{FigProfile0742}As in Figure \ref{FigProfile0631}, but
for Parkes data of PSR J0742--2822 at 1369 MHz.
}
\end{figure}

Like the other pulsars discussed so far, the radio emission of this
pulsar is highly linearly polarized, but there is
depolarization at the trailing edge (see Figure \ref{FigProfile0742}).
The circular polarization changes sign roughly in the middle of the
profile, which is often associated with emission coming from close to
the magnetic axis \citep{rr90}.

\begin{figure}[t]
\epsscale{.70}
\begin{center}
\includegraphics[angle=270,width=\figwidth]{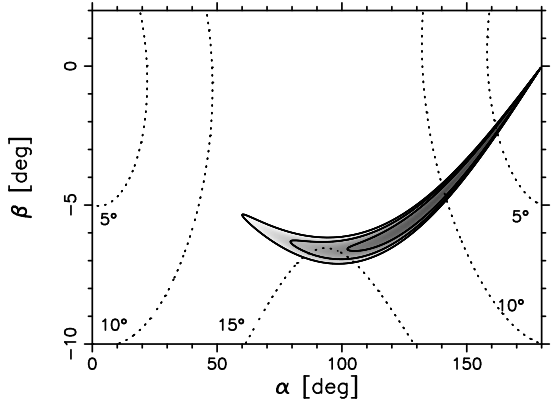}
\end{center}
\caption{\label{FigRVM0742}As in Figure \ref{FigRVM0631}, but for PSR
J0742--2822. The lowest reduced $\chi^2$ is 6.9.
}
\end{figure}

The $\chi^2$ map of the RVM fit (Figure \ref{FigRVM0742}) is much
better constrained than for the other pulsars. The reduced $\chi^2$ of
the best fit is not good (6.9), because the observed
PA-swing deviates slightly from the model at the far trailing side of
the profile (see Figure \ref{FigProfile0742}). This deviation is
caused by a small jump in PA at the pulse phase corresponding to
the pulse phase where significant depolarization is observed.

\cite{wj08b} derived that $h_\mathrm{PA}=350$ km, based on the
observed offset between the center of the radio pulse profile 
and the location of the steepest gradient of the PA-swing
(slightly larger than that derived by e.g. \citealt{hx97}),
suggesting $\rho=\degrees{18}$ (Equation \ref{EqRho}). This opening
angle is inconsistent with the RVM fit (see Figure \ref{FigRVM0742}),
which suggests that the half opening angle of the radio beam
$\rho\lesssim\degrees{15}$ (in line with the value of $\rho$
derived by e.g. \citealt{kwj+94}).  Possibly the beam is asymmetric
with respect to the magnetic axis (e.g. \citealt{lm88}).
Nevertheless, if we believe that most of the open field line region is
active, this pulsar is unlikely to be aligned and we expect
$\beta\sim\degrees{-7}$.

\subsection{PSR J1420--6048}

\begin{figure}[t!]
\begin{center}
\includegraphics[angle=0,width=\figwidth, trim=30 0 50 0, clip=true]{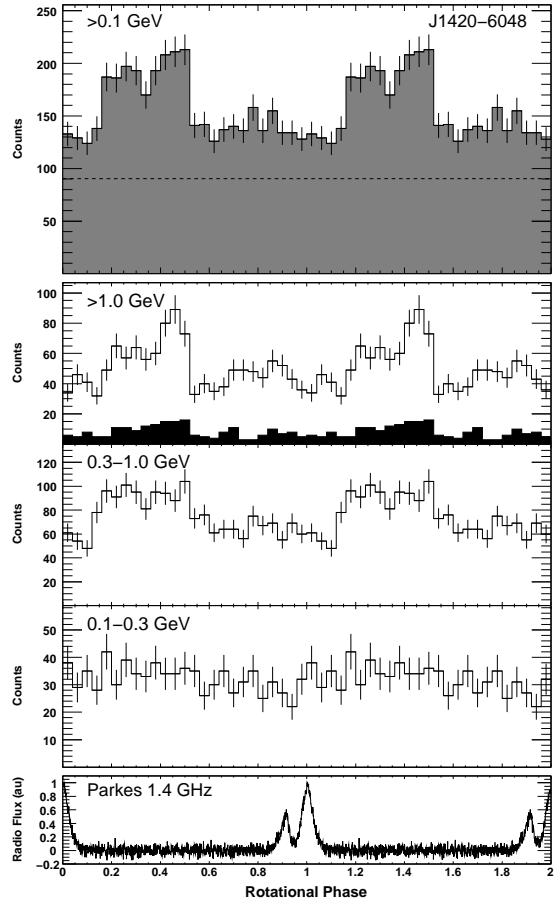}
\end{center}
\caption{\label{Fig1420}As in Figure \ref{Fig0631}, but for PSR J1420--6048.
}
\end{figure}

\subsubsection{The pulsar and its surroundings}

\object{PSR J1420--6048} is a 68 ms pulsar \citep{dkm+01} in the
northeast wing of the complex of compact and extended radio sources
known as Kookaburra \citep{rrj01}. It is located within the 0\fdg32 wide 95\%
confidence level radius of the center of 3EG J1420--6038.
The pulsar's large spin-down power and distance makes it a
plausible match for the EGRET source. The situation is complicated by the {\em Fermi} LAT discovery of
pulsations from an X-ray source in the Rabbit pulsar wind nebula,
PSR J1418--6058 with $\dot E = 5\times 10^{36}$ erg\,s$^{-1}$, only
0\fdg24 away from the radio pulsar, and 0\fdg54 from the
EGRET source \citep{aaa+09c}.
The period is 110 ms,
different from the weak detection of pulsed X-rays with period 108 ms
reported by \cite{nrr05}. 
There are also three TeV sources in this region discovered by HESS
\citep{aab+06b,aab+08}, but the association between the TeV sources
and the $\gamma$-ray pulsars is unclear.

\object{PSR J1420--6048} is the youngest and most energetic pulsar of
our sample. The pulsar distance, derived from the DM to be $5.6\pm0.9$
kpc according to the \cite{cl02} model, is consistent with the X-ray
absorption of $N_\mathrm{H}\sim2\times10^{22}$ cm$^{-2}$ \citep{rrj01}
but subject to uncertainties because of {\sc HII} regions and dense
clouds in the Carina arm. In this paper we apply a more
conservative distance uncertainty of 30\% to our adopted DM
distance estimates (so $5.6\pm1.9$ kpc for this pulsar). This is to
try to take into account systematic errors in the model for the
electron density in the Galaxy and this approach is identical to that
in the LAT pulsar catalog paper \citep{aaa+09e}.

\subsubsection{$\gamma$-rays}

PSR J1420--6048 is the brightest $\gamma$-ray pulsar of our
sample. The light curve of PSR J1420--6048 (see Figure \ref{Fig1420})
has a broad peak, which probably consists of two components lagging
the second radio peak by $0.26$ and $0.44$ in phase. This
would imply a peak separation $\Delta=0.18$.  The second component may
follow the common {\it Fermi} pulsar pattern of increasing dominance
at high $\gamma$-ray energies \citep{aaa+09e}.
The small angular separation between this pulsar and J1418--6058
\citep{aaa+09c} considerably increases the background
flux at the position of the pulsar over a large energy
interval. The estimation of the background level, which follows from a
simple measurement of the flux in a \degrees{1}--\degrees{2} ring around the
pulsar, does not take into account the flux of J1418--6058, and as a
consequence the background level is underestimated. 
The $\gamma$-ray spectrum is fitted significantly better by
including an exponential cutoff.

\subsubsection{X-rays}

\begin{figure}[t]
\begin{center}
\includegraphics[angle=270,width=\figwidth]{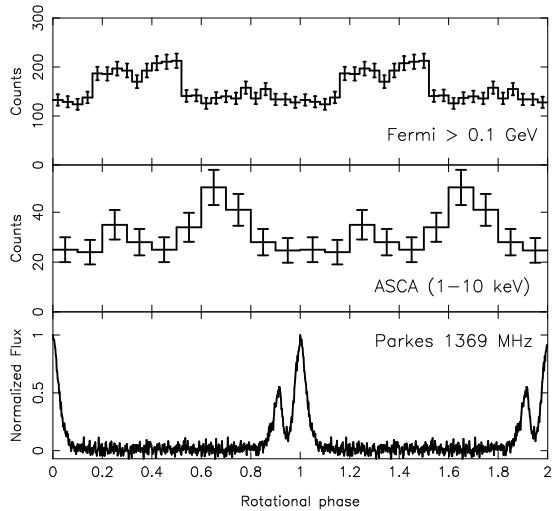}
\end{center}
\caption{\label{FigMulti1420}Compilation of phase-aligned light curves
of PSR J1420--6048 as seen by {\em Fermi}, ASCA \citep{rrj01} and
Parkes. The X-ray light curve has an absolute phase error
$\sim0.06$. }
\end{figure}

There is a marginal detection of X-ray pulsations at the radio pulse
period by {\em ASCA} (see Figure \ref{FigMulti1420}) from within the
X-ray nebula AX J1420.1--6049 \citep{rrj01}. 
The X-ray and $\gamma$-ray light curves peak at different phases
with respect to the radio peak, so it is not clear how the
$\gamma$-rays and X-rays are related.

\subsubsection{Radio}

\begin{figure}[t]
\epsscale{.70}
\begin{center}
\includegraphics[angle=270,width=\figwidth]{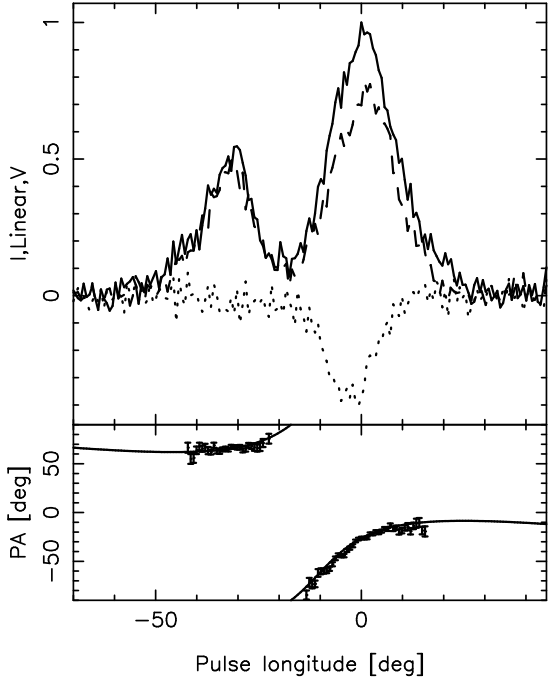}
\end{center}
\caption{\label{FigProfile1420}
As in Figure \ref{FigProfile0631}, but
for Parkes data of PSR J1420--6048 at 1369 MHz.
}
\end{figure}

PSR J1420--6048 is another example of a radio pulsar that is nearly
completely linearly polarized (e.g. \citealt{jw06}) and there is also
some circular polarization (see Figure \ref{FigProfile1420}). The
radio pulse profile shows a double-peaked structure with the trailing
component being strongest, something that is generally seen for young
pulsars with characteristic ages less than 75 kyr \citep{jw06}.

\begin{figure}[t]
\epsscale{.70}
\begin{center}
\includegraphics[angle=270,width=\figwidth]{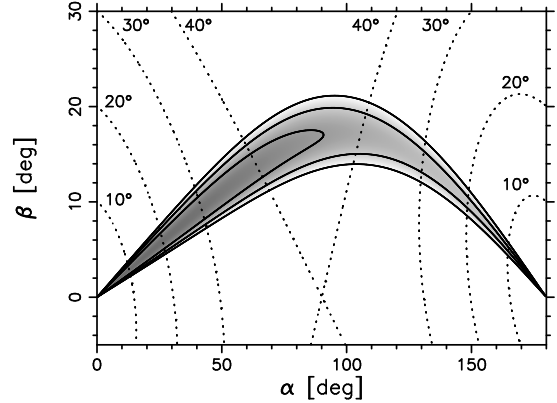}
\end{center}
\caption{\label{FigRVM1420}As in Figure \ref{FigRVM0631}, but for PSR
J1420--6048. The lowest reduced $\chi^2$ is 1.3.
}
\end{figure}

RVM modelling of the PA-swing has been carried out by \cite{rrj01} who
claim $\alpha\le\degrees{35}$ and $\beta\sim0\fdg5$. It is
clear from Figure \ref{FigRVM1420} that the best solution can be found
at low $\alpha$ values, but neither $\alpha$ nor $\beta$ are well
constrained.
This geometry would be consistent with the arc of emission around the
pulsar seen in the {\em Chandra} image \citep{nrr05}, which suggests (if
interpreted as a torus) a $\zeta$ which is not particularly small or
large.

\cite{wj08b} found that $h_\mathrm{PA}=100$ km, which is of the same
order as what was found by \cite{jw06} (175 km). Using the emission
height of 100 km in Equation \ref{EqRho} leads to an expected $\rho$
of \degrees{15}. This implies (see Figure \ref{FigRVM1420}) that the
magnetic inclination angle $\alpha$ should be relatively small
($\sim\degrees{20}$). As always, this derivation relies on the
assumption that the pulsar beam is symmetric and nearly filled.

PSR J1420--6048 belongs to a group of young pulsars with very wide
profiles and relatively high values of $\dot{E}\gtrsim5\times10^{35}$
erg\,s$^{-1}$ (e.g. \citealt{man05,wj08b}). They speculate that an
analogue can be drawn between the radio emission of these so-called
``energetic wide beam pulsars'' and their high-energy emission.  They
argued that the sites in the pulsar magnetosphere that produce the
radio emission could be very similar to those of the high-energy
emission, leading to the prediction that there should be strong
similarity between the radio and $\gamma$-ray light curves. Although
the $\gamma$-ray light curve may indeed be double peaked, the
$\gamma$-ray light curve is significantly offset in phase from the
radio profile, suggesting a significant difference in the location of
production of the radio and $\gamma$-ray emission.

\subsection{PSR J1509--5850}

\begin{figure}[t!]
\begin{center}
\includegraphics[angle=0,width=\figwidth, trim=30 0 50 0, clip=true]{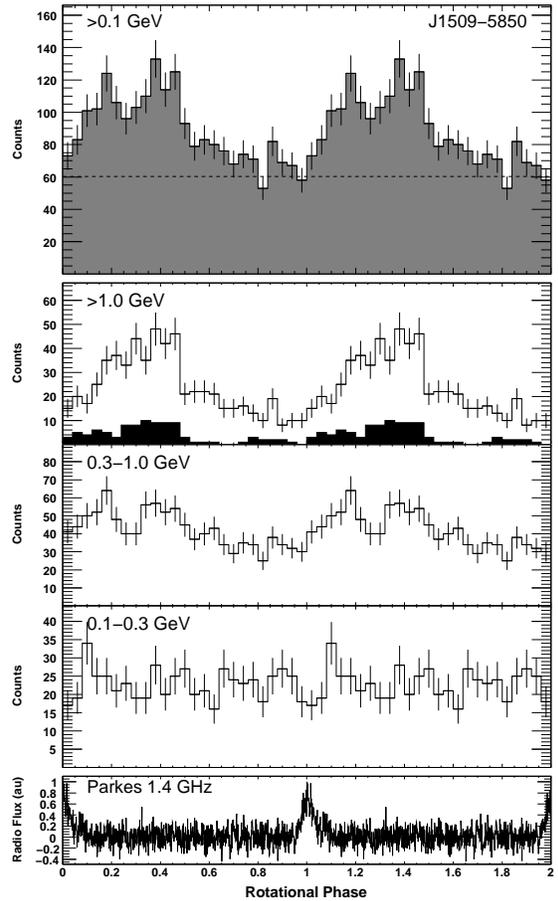}
\end{center}
\caption{\label{Fig1509}As in Figure \ref{Fig0631}, but for PSR J1509--5850.
}
\end{figure}

\subsubsection{The pulsar and its surroundings}

\object{PSR J1509--5850} was discovered as a 89 ms radio pulsar by
\cite{kbm+03}. It has a pulsar wind nebula as well as a long tail seen
in X-rays (e.g. \citealt{kmp+08}) and radio \citep{hb07}.  The
distance to this pulsar estimated from the DM is $2.5\pm0.5$ kpc
using the \cite{cl02} model, but, as discussed before, we adopt
a more conservative distance uncertainty of 30\% ($2.5\pm0.8$ kpc).

\subsubsection{$\gamma$-rays}

\begin{figure}[t]
\epsscale{.70}
\begin{center}
\includegraphics[angle=270,width=\figwidth]{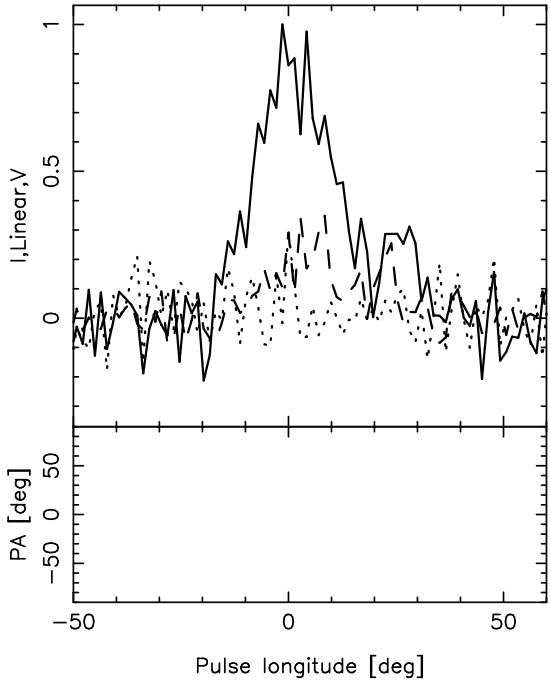}
\end{center}
\caption{\label{FigProfile1509}As in Figure \ref{FigProfile0631}, but
for Parkes data of PSR J1509--5850 at 1369 MHz.
}
\end{figure}

This pulsar should not be confused with PSR B1509--58 (PSR
J1513--5908), which was seen in soft $\gamma$-rays by BATSE, OSSE, and
COMPTEL on {\em CGRO} \citep{umw+93}. The light curve of PSR J1509--5850
(see Figure \ref{Fig1509}) is very broad. Similarly to PSR J1420--6048,
the light curve may be composed of two peaks lagging the radio profile
by $0.18$ and $0.39$ in phase respectively (corresponding to a peak separation $\Delta=0.21$).
The spectrum of this pulsar shows a cutoff at $\sim3$ GeV with
the highest confidence of the pulsars discussed in this paper.
The derived $\gamma$-ray efficiency is very high ($\sim15\%$) and
is the largest of the pulsars discussed in this paper (see Table
\ref{tableGammaLum}). However, the distance to this pulsar is highly
uncertain, therefore the luminosity and hence the $\gamma$-ray
efficiency are not well constrained.

\subsubsection{Radio}

\begin{figure}[t!]
\begin{center}
\includegraphics[angle=0,width=\figwidth, trim=30 0 50 0, clip=true]{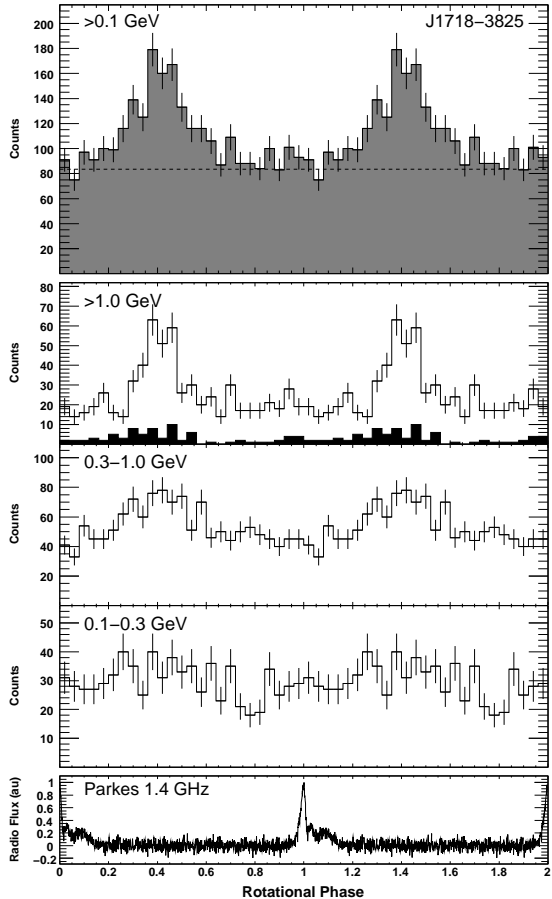}
\end{center}
\caption{\label{Fig1718}As in Figure \ref{Fig0631}, but for PSR J1718--3825.
}
\end{figure}

This pulsar is the weakest radio source of our sample. As noted by
\cite{wj08b}, the correlation between $\dot{E}$ and a high degree of
linear polarization does not hold for this pulsar (see Figure
\ref{FigProfile1509}). The low degree of linear
polarization
in combination with a low overall signal-to-noise ratio prevents us from
measuring the PA-swing, and therefore the emission geometry cannot be
constrained for this pulsar using the RVM model.

\subsection{PSR J1718--3825}

\begin{figure}[t]
\epsscale{.70}
\begin{center}
\includegraphics[angle=270,width=\figwidth]{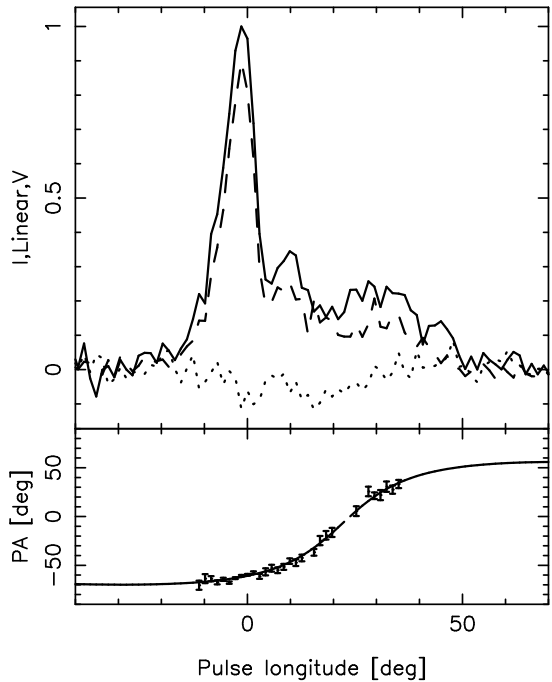}
\end{center}
\caption{\label{FigProfile1718} As in Figure \ref{FigProfile0631}, but
for Parkes data of PSR J1718--3825 at 1369 MHz.
}
\end{figure}

\subsubsection{The pulsar and its surroundings}

\object{PSR J1718--3825} is a 75 ms pulsar discovered by
\cite{mlc+01} at radio wavelengths. It has an associated X-ray nebula
\citep{hfc+07} and an associated HESS source \citep{aab+07}. Its
distance derived from the DM
is $3.6\pm0.4$ kpc according to the \cite{cl02} model, or $3.6\pm1.1$
kpc by applying a more conservative distance uncertainty of 30\%. 

\subsubsection{$\gamma$-rays}

The light curve of PSR J1718--3825 is single peaked (FWHM is $0.20$ in
phase) and it lags the radio profile by $0.42$ in phase (see Figure
\ref{Fig1718}). Like the previous two pulsars discussed its spectrum is
significantly better described by including a cutoff energy. However,
the cutoff energy itself cannot be determined well.

\subsubsection{Radio}

\begin{figure}[t]
\epsscale{.70}
\begin{center}
\includegraphics[angle=270,width=\figwidth]{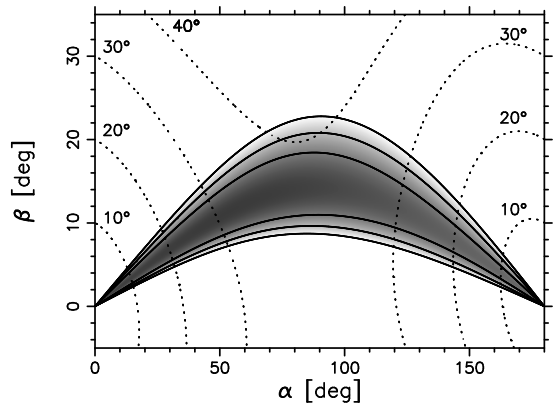}
\end{center}
\caption{\label{FigRVM1718}As in Figure \ref{FigRVM0631}, but for PSR
J1718--3825. The lowest reduced $\chi^2$ is 1.3.
}
\end{figure}

The radio pulse profile of J1718--3825 has a relatively complex shape
and it is highly linearly polarized (see Figure
\ref{FigProfile1718}). There is some negative circular polarization as
well. The 1.4 GHz profile is very similar to what is seen at higher
frequencies (unpublished Parkes data\footnotemark[4]).

The PA-swing is S-shaped, allowing an RVM fit, although it is not
very constraining (see Figure \ref{FigRVM1718}). The steepest gradient
of the RVM model lags the center of the radio pulse by $\degrees{5}$,
corresponding to $h_\mathrm{PA}\sim80$ km (Equation \ref{EqBCW}) and an
expected $\rho\sim\degrees{13}$ (Equation \ref{EqRho}). This would
imply that $\alpha\sim\degrees{20}$ (see Figure \ref{FigRVM1718}),
however, the complex shape of the radio profile makes it difficult to
objectively determine which pulse phase corresponds to emission from
the fiducial plane. If the peak of the profile is a better indicator
for the magnetic axis than the center, then $\alpha$ is larger
($\sim\degrees{50}$).

\section{Discussion}
\label{SctDiscussion}

From examination of the light curves in the LAT pulsar catalog
\citep{aaa+09e}, certain groupings of $\gamma$-ray light curves are
apparent. One group, of which the Vela pulsar is the archetypical
example \citep{aaa+09d}, is characterized by light curves which consist of
two narrow peaks separated by $\sim$0.4--0.5 in phase with the first
peak offset from the radio profile by $\sim$0.1--0.2 in phase. A
second group of light curves also shows double peaks, but the peaks
are much closer together in phase and can blend together to produce a
more square-looking light curve. Examples of this sort include PSR
J1709--4429, the radio-quiet PSR J0007+7303 \citep{aaa+09c} and the radio loud PSR
J1057--5226 \citep{aaa+09e}. The final group of light curves appear to
consist of a single component, possibly because the peaks are
completely blended together, are unresolved given the present
signal-to-noise ratio or simply because they are truly single.
There is some evidence that the
closer doubles and single component light curves are found
preferentially among the pulsar with lower $\dot{E}$, whereas the
Vela-like light curves are found for all spin-down luminosities
\citep{aaa+09e}.

The light curves of the six pulsars discussed in this paper are
all consistent with single peaks.  Nevertheless,
those of PSRs J1420--6048 and J1509--5850 could have the appearance of
closely-spaced doubles,
therefore resembling that of PSR~J1709--4429. The light curves of the
other four pulsars do not show,  at least
given the current count statistics,
any hint of a
second pulse component. PSRs~J0631+1036 and J1718--3825 have
relatively broad pulses with an offset to the radio pulse of 0.4
in phase. The $\gamma$-ray emission of PSR J0631+1036 has a tail to
earlier phase, visible especially above 1 GeV. The light curve of
PSR~J0742--2822 is narrower, and is offset from the radio peak by 0.61
in phase. Finally, PSR~J0659+1414 has a single broad peak at the same
phase as a corresponding peak in the optical and UV and yet lacks the
second peak seen at those energies.

We now consider if outer magnetospheric models can explain these light
curves in conjunction with the constraints on the geometry 
derived from the radio profiles by comparing the observed $\gamma$-ray
light curves to those predicted in the two-pole caustic and outer-gap
models of \cite{wrwj09}.

In the context of the \cite{wrwj09} models the light curves of PSRs J0631+1036,
J1420--6048, J1509--5850 and J1718--3825 can all be described in term
of closely-spaced double peaks. For PSRs J1420--6048 and J1509--5850
there is some evidence for a double-peaked nature, while for
PSRs~J0631+1036 and J1718--3825 it is well possible that the peaks are
unresolved due to a lack of signal-to-noise ratio. 
Double-peaked light curves with small separations occur naturally at the correct phases relative to
the radio emission in the outer-gap and two-pole caustic pictures, if
one skims the edge of the outer magnetosphere cone. For narrow gaps
(low-efficiency pulsars) this occurs for a range $\alpha\sim
40-50^\circ$ and $\zeta \sim 50-65^\circ$ for the outer-gap model and
$\alpha \sim 45-60^\circ$ and $\zeta \sim 35-55^\circ$ in the two-pole
caustic model. For wider gaps (high-efficiency pulsars), the outer-gap
model has many similar solutions extending to $\alpha, \zeta \sim
80^\circ$, but such pulses are more difficult to realize in the
two-pole caustic geometry.

Single peaks that lag the radio pulse by 0.4 -- 0.5 in phase occur in
both outer-gap and two-pole caustic models for the smallest gap
widths, over a band spanning $\alpha \approx 55-75^\circ$, $\zeta
\approx 20-50^\circ$.  These are effectively the second peak of the
more typical double profile, with the first peak weak or absent in
this angle range. 
The preferred geometry derived from the
radio data of PSR J1420--6048 ($\alpha\sim\degrees{20}$ and
$\beta\sim\degrees{5}$) would favour the two-pole caustic model (the
outer-gap model does not predict significant $\gamma$-ray pulsations
for such a geometry). However, in this case we would expect the
$\gamma$-ray peak to appear at phase $0.1-0.2$. In any case, at
present the constraints derived from the radio data are too weak to
draw firm conclusions.

If the light curve of PSR~J0742--2822 consists only of a single narrow
peak at phase 0.61 it cannot easily be explained by either the
outer-gap or two-pole caustic model. However, it is tempting to
associate this peak with the second peak seen in the Vela-like light
curves as it has the correct phase offset with respect to the radio
profile. If we do this then the first peak must either be missing
entirely or at least be much weaker than the second peak; we note that
the weak excess at phase $0.12$ in Figure \ref{Fig0742} is at the
expected phase -- longer integrations should eventually settle
this. This picture would tie in nicely with the general trend of a
relatively strong second peak at higher energies as seen for double-peaked light curves.

Finally we discuss the curious case of PSR J0659+1414. It has a high
signal-to-noise ratio light curve and is clearly single. The spin
parameters of PSR J0659+1414 are not exceptional. Its $\dot{E}$ is
relatively low, but that of PSR J1057--5226, which has a harder
spectrum, is lower. However, the magnetic field strength at the light
cylinder is very low for J0659+1414. A special geometry may be
responsible for the unusual properties of this pulsar and we consider
two possible interpretations.
In the first, PSR~J0659+1414 is more Vela-like. In this scenario, the
second component (visible at lower energies) is entirely missing at
$\gamma$-ray energies. However, although the phase of the visible
component can then be well explained by both outer-gap and two-pole
caustic models, it implies relatively large values for $\alpha$ and
$\zeta$ which is somewhat at odds with the constraints from the
modelling of the radio data and the thermal X-rays. In the second
interpretation PSR~J0659+1414 is an aligned rotator
($\alpha\lesssim\degrees{40}$). In this case in the outer-gap picture
one would not expect strong emission, but the two-pole caustic model
does indeed show a single pulse at phase $0.1-0.2$ later than the
radio emission. We also note that the pulsar has an extremely low
efficiency (less than $\sim$1\%) and as its distance is well known and
the flux correction factor $f_\Omega$ is likely to be of order unity
there is little way to avoid this conclusion.  Further, its spectrum
is among the softest of all the pulsars detected by {\em Fermi}
\citep{aaa+09e} and it is virtually undetected at energies $>1$
GeV. This might also point to a special, aligned geometry where
$\gamma$-ray emission at somewhat lower altitude ($\ga
0.1R_\mathrm{LC}$) is being observed.

In summary, therefore, we find that both the two-pole caustic model
and the outer-gap model do a good job in predicting the phase offset
with respect to the radio emission for PSRs J0631+1036, J1420--6048,
J1509--5850 and J1718--3825 under the assumption that these are
closely-spaced double-peaked $\gamma$-ray emitters. We note that the
light curve shape, especially of PSR~J1718--3825, is not well
predicted by the models. However, it should be stressed that the model
calculations in \cite{wrwj09} are idealized and therefore one would
indeed only expect approximate agreement between the models and the
data.  One could speculate that to obtain more realistic physical
models one should include a larger range of field lines from which  $\gamma$-ray photons are produced. This smooths out the sharp peaks in the
light curves caused by caustics; these sharp peaks are not observed for the
pulsars in this paper. This also has the effect of making the so-called
``bridge'' of emission between the peaks stronger, causing the
peaks in the light curve to blended together. In \cite{wrwj09} the
two-pole caustic and outer-gap widths increase with $\dot{E}$, but the
emission region is an infinitely thin strip on the gap inner edge.
For PSR~J0742--2822, we surmise that there is a ``missing'' $\gamma$-ray
peak at phase offset 0.1 (and possibly the second peak is
``missing'' for PSR J0659+1414). Without speculating about a missing
peak, it is hard to understand the light curve of PSR~J0659+1414 in
the context of outer-gap models. It may be a roughly aligned rotator
with $\gamma$-ray emission from lower down in the magnetosphere.

\section{Conclusions}
\label{SctConclusions}

We report here on the detection of pulsed $\gamma$-rays by {\em Fermi}
for PSRs J0631+1036, J0659+1414, J0742--2822, J1420--6048, J1509--5850
and J1718--3825.  These six pulsars are young to middle-aged and,
except for PSR J1420--6048, have relatively small values of $\dot{E}$
compared to other known $\gamma$-ray pulsars. In all cases the
$\gamma$-ray light curves appear single peaked (at least with the present
count statistics), but there is a hint that at least two of them have
closely-spaced double peaks. As {\em Fermi} continues its all-sky
survey, the quality of the light curves will increase, helping to
resolve this issue.

We present high quality radio polarization profiles for these pulsars
and discuss their geometries in the context of RVM fitting and simple
beam modelling. Unfortunately, the narrow phase range of the radio
emission generally leaves a strong degeneracy between the fit $\alpha$
and $\beta$ values. This, in combination with the
limited $\gamma$-ray count statistics makes it difficult to distinguish between
single-peaked and double-peaked light curves and hence to make the
comparison with the model predictions.  We show that models where the
$\gamma$-ray emission occurs at relatively high altitudes in the
pulsar magnetosphere, such as the outer-gap or two-pole caustic
models, do a good job in predicting the phase of the $\gamma$-ray
emission relative to the radio emission.  However, the shape of the
$\gamma$-ray light curve is less well modelled, and additional inputs
to the model are needed to explain the strong bridge emission in
some pulsars and the strong single $\gamma$-ray component in
PSR~J0742--2822 in particular. Finally, PSR~J0659+1414 warrants
further attention.  It has a peculiar light curve and phase offset
with the radio profile, its $\gamma$-ray efficiency is low and its
$\gamma$-ray spectrum is extremely soft. It may be an aligned rotator
with $\gamma$-ray emission arising relatively low down in the
magnetosphere. Higher signal-to-noise ratio $\gamma$-ray light curves
in combination with possible additional geometrical constraints, such
as from PWN imaging, will result in stronger constraints on the
models.



\acknowledgments

The Australia Telescope is funded by the Commonwealth of Australia for
operation as a National Facility managed by the CSIRO.
The Nan\c cay Radio Observatory is operated by the Paris Observatory,
associated with the French Centre National de la Recherche Scientifique (CNRS).
The Lovell Telescope is owned and operated by the University of Manchester as
part of the Jodrell Bank Centre for Astrophysics with support from the Science
and Technology Facilities Council of the United Kingdom.

The $Fermi$ LAT Collaboration acknowledges generous ongoing support from a
number of agencies and institutes that have supported both the development and
the operation of the LAT as well as scientific data analysis.  These include
the National Aeronautics and Space Administration and the Department of Energy
in the United States, the Commissariat \`a l'Energie Atomique and the Centre
National de la Recherche Scientifique / Institut National de Physique
Nucl\'eaire et de Physique des Particules in France, the Agenzia Spaziale
Italiana and the Istituto Nazionale di Fisica Nucleare in Italy, the Ministry
of Education, Culture, Sports, Science and Technology (MEXT), High Energy
Accelerator Research Organization (KEK) and Japan Aerospace Exploration Agency
(JAXA) in Japan, and the K.~A.~Wallenberg Foundation, the Swedish Research
Council and the Swedish National Space Board in Sweden.

Additional support for science analysis during the operations phase is
gratefully acknowledged from the Istituto Nazionale di Astrofisica in
Italy and the Centre National d'\'Etudes Spatiales in France.






\bibliographystyle{apj}

\end{document}